\documentstyle[prl,aps,epsf]{revtex}
\draft
\begin{document}
\twocolumn[\hsize\textwidth\columnwidth\hsize\csname @twocolumnfalse\endcsname

\title{Hysteresis and Noise in   
Stripe and Clump Forming Systems} 
\author{C. Reichhardt, C.J. Olson Reichhardt, and A.R.~Bishop} 
\address{ 
Center for Nonlinear Studies and 
Theoretical Division, 
Los Alamos National Laboratory, Los Alamos, New Mexico 87545}
\date{\today}
\maketitle

\begin{abstract}
We use simulations to examine hysteresis and noise in 
a model system that produces
heterogeneous orderings including
stripe and clump phases. 
In the presence of a disordered substrate,
these heterogeneous phases exhibit $1/f^{\alpha}$ 
noise and 
hysteresis in transport. The noise 
fluctuations are maximal in the heterogeneous phases,  while in the uniform  
phases the hysteresis
vanishes and both $\alpha$ and the noise power decrease.   
We compare our results to recent experiments 
exhibiting noise and hysteresis
in high-temperature superconductors 
where charge heterogeneities may occur.
\end{abstract}


\vskip2pc]
\narrowtext
There is growing evidence that a wide variety 
of condensed matter systems 
intrinsically exhibit heterogeneous charge ordering in 
the form of mesoscopic clump, labyrinth, or stripe phases. 
Examples of systems where
such phenomena may occur include cuprate superconductors
\cite{Gorkov,Zaanen,Emery,StripeReview1,Bishop2}, 
antiferromagnetic insulators \cite{Cho}, and
two-dimensional electron gas (2DEG) systems 
\cite{Electron3,Kivelson4}. 
Heterogeneous states are observed in 
manganites and diluted magnetic semiconductors \cite{Dagotto5}
as well as superconductors \cite{Tranquada,Davis}.  
In most of these systems, some form of quenched disorder 
from intrinsic defects in the sample is present, and it
can destroy any long range ordering in the patterns
\cite{Vlad2}. 
Charged stripe forming systems have also been shown to exhibit 
self-generating disordered glassy properties 
\cite{Schmalian6,Schmalian2,Schmalian6b}. 
The presence of ordered or disordered heterogeneities should affect 
the bulk transport, fluctuations,
and transient responses of the systems; however, little is known about 
how the effect of heterogeneity on transport would differ from 
that of homogeneous but disordered systems which form uniform
crystalline or partially crystalline phases. 

Recently, transport experiments in underdoped samples of YBCO have 
revealed
hysteretic jumps in the current-voltage curves 
at low temperatures, as well as 
non-Gaussian noise fluctuations
in the resistance curves \cite{Bonetti7}. 
As the temperature is increased,
the current-voltage curves become smooth and non-hysteretic.     
These results have been interpreted as a signature 
of some form of large scale heterogeneities
such as disordered fluctuating domains.
Other experiments on cuprate superconductors
in the non-superconducting region of the phase diagram
have shown magnetic hysteresis and avalanchelike jumps
in the magnetization curves \cite{Pan8,Pan29}, 
reminiscent of the Barkhausen noise that
occurs for domain wall depinning in 
ferromagnets. 
This magnetic hysteresis vanishes
at low and high doping and at high temperatures \cite{Pan29}.
Transport experiments in 2DEGs have
also produced hysteretic current-voltage 
curves which are believed to indicate the formation of pinned 
clump or bubble phases \cite{Cooper10}.   

In this work we study the noise fluctuations and hysteresis in the 
transport curves for a two-dimensional model that exhibits uniform, 
clump, and disordered stripe phases as a function of density. 
Our model consists of interacting particles with a repulsive
Coulomb interaction and an additional short range exponential attraction.
At zero temperature and in the absence of quenched disorder,
a uniform crystal occurs at low density. As the density
is increased the system forms a clump phase, then a stripe or labyrinth phase
followed by an anti-clump phase and finally a uniform partially 
crystalline phase at high densities. 
When quenched disorder is present the phases are more disordered 
and can be pinned in the presence of a driving force. 

We find that the heterogeneous clump and stripe phases 
exhibit hysteresis in the transport curves accompanied by a 
large increase in the noise
power with a $1/f^{\alpha}$ noise spectrum characteristic, 
where $\alpha > 1$. In the uniform phase
at low and high densities, the hysteresis vanishes and 
$1/f$ noise appears.
As the temperature is increased in the heterogeneous phases, 
the hysteresis is lost and the transport curves become smooth
with an accompanying decrease in the noise fluctuations. We find that the
hysteresis occurs in a dome-shaped region of the temperature and density 
phase diagram. 
Importantly, these properties require {\it both} the intrinsic
heterogeneity and the pinning mechanisms.
Our results are consistent with the clump and stripe phases forming a glassy
or domain glass phase.  

{\it Model}---We consider a two-dimensional system
with periodic boundary conditions in the $x$ and $y$ directions. 
We fix the system size to $90 \times 90$ 
and vary the particle density $n$ by changing
the number of particles $N$ from 80 to 3000.
The  particle-particle interaction 
${\bf f}_{ij}=-\nabla U(r_{ij}){\bf \hat r}_{ij}$ 
for particles separated by ${\bf r}_{ij}={\bf r}_i-{\bf r}_j$
consists of a long-range 
Coulomb repulsion and a short range exponential attraction: 
\begin{equation}
U(r) = 1/r - B\exp(-\kappa r).
\end{equation}
At small and large $r$ the repulsive 
Coulomb term dominates. The attractive interaction 
can be varied using 

\begin{figure}
\center{
\epsfxsize=3.5in
\epsfbox{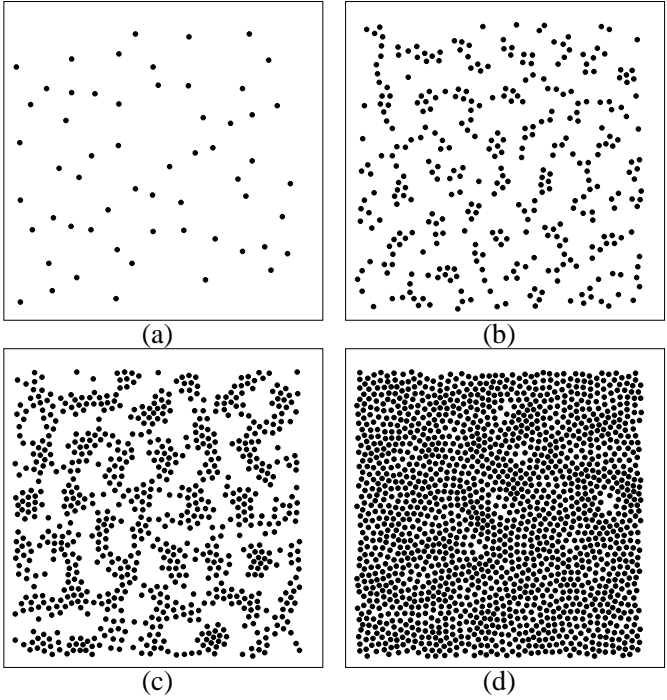}}
\caption{
The particle positions (black dots) after annealing to $T = 0$ for
densities (a) $n = 0.013$, (b) $ n = 0.06$, (c) $n = 0.15$, and 
(d) $n = 0.34$.           
}
\end{figure}

\noindent
the inverse screening length 
$\kappa$ and the parameter $B$;
in this work both $B$ and $\kappa$ are held fixed.  
In previous work we have shown that this model
produces crystal, clump, stripe, and anti-clump phases as $B$ and the
density $n$ are varied \cite{Reichhardt11}. 
Here we also consider the effects of quenched disorder,  
modeled as $N_p$
randomly placed attractive parabolic pins of
strength $f_{p}=2.5$ and radius $r_{p}$,
giving ${\bf f}_p=\sum_{k=1}^{N_p}(f_p/r_p){\bf r}_{ik} 
\Theta(1-r_{ik}/r_p)$, 
where $\Theta$ is the Heaviside step function,
for particle-pin spacing of ${\bf r}_{ik}={\bf r}_i-{\bf r}_k$.
The overdamped equation of motion for a single particle $i$ is 
$\eta {\bf V}_i = {\bf f}_{i} 
= \sum_{j\ne i}^N{\bf f}_{ij} + {\bf f}_{p} + {\bf f}_{d} + {\bf f}^{T}.$     
Here $\eta$ is a phenomenological damping term.   
The driving term ${\bf f}_{d}=f_d{\bf \hat{x}}$ would come from 
an applied voltage in the case of charged particles. 
The conduction is proportional to the  
average particle velocity $<V>$ in the direction
of the applied drive.  We increase the drive
in small increments and average over many thousands of time steps
to avoid transient effects and ensure the velocities have
reached a steady state. 
The temperature is modeled as Langevin random kicks with the properties
$<f^{T}(t)> = 0$ and $<f^{T}(t)f^{T}(t^{\prime})> = 
2\eta k_{B}T\delta(t-t^{\prime})$.
The system is initially prepared in a high temperature molten state
and annealed down to a lower finite temperature, after which
the driving force is applied. The parameters we vary in this work are
the density, driving force, and temperature. 
We have performed a similar series of simulations with $f_p=1.5$ and
find the same behavior at shifted values of drive.  We have also 
checked the effect of system size using a 40 $\times$ 40 system, and
find the same results shown 

\begin{figure}
\center{
\epsfxsize=3.5in
\epsfbox{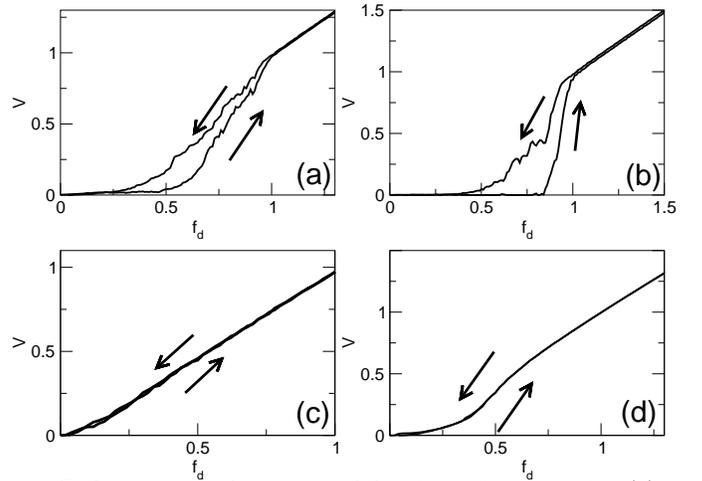}}
\caption{
The velocity $V$ vs driving force $f_d$ curves for 
(a) density $n = 0.15$, $T = 0.1$; 
(b) $n = 0.06$, $T = 0.1$; 
(c) $n = 0.15$, $T = 1.3$;
(d) $n = 0.34$, $T=0.1$.
}
\end{figure}

\noindent
here but with lower resolution.

In Fig.~1 we illustrate 
some of the representative phases that occur with increasing 
density in the presence of the
quenched disorder. In Fig.~1(a) at density $n = 0.013$, 
the particles form a uniform phase of single charges. 
In Fig.~1(b) at $n = 0.06$, the system 
is comprised of heterogeneous arrangements of disordered clumps. 
For higher density, such as $n = 0.15$ shown in Fig.~1(c), the system forms 
a disordered labyrinth pattern.  At very high densities 
the system returns to a uniform phase 
with considerable crystalline order as seen
in Fig.~1(d) for $n = 0.34$. 
The heterogeneous phases occur for $ 0.05 < n < 0.34$.    

By examining the
velocity-force curves at $T = 0$, we find smooth non-hysteretic
curves for densities where uniform phases occur,
while in the heterogeneous regions, $V(f_d)$ shows
pronounced hysteresis. In Fig.~2(a) we plot the velocity vs applied
force curve for a heterogeneous system at $n = 0.15$ where the equilibrium
state is a labyrinth phase. 
Here the curve shows two abrupt changes in slope,
and $V$ on the decreasing sweep of $f_d$ is higher than on the increasing
sweep.  If the drive is increased again the same ramp up curve is  
followed. 
In Fig.~2(b),
$V(f_d)$ is shown 
for a system with $n=0.06$, 
where the equilibrium system forms a clump phase. Here
a hysteretic region is also observed. 
The current-voltage curve for $n=0.34$, plotted in
Fig.~2(d),
is smooth and has no hysteresis within our resolution.
This density $n=0.34$ corresponds to the uniform phase seen in 
Fig.~1(d). 
For the densities that exhibit hysteresis, 
as the temperature is 
increased the width of the hysteresis is reduced and 
it vanishes completely at high temperatures. 
In Fig.~2(c) 
we plot the velocity-force curve for the system in 
Fig.~2(a) for a temperature of $T=1.3$ 
where the hysteresis has been lost and the 
velocity force curve is smooth.  

In order to better characterize 
where the heterogeneous phases occur 
and where a hysteretic response is observed, 

\begin{figure}
\center{
\epsfxsize=3.5in
\epsfbox{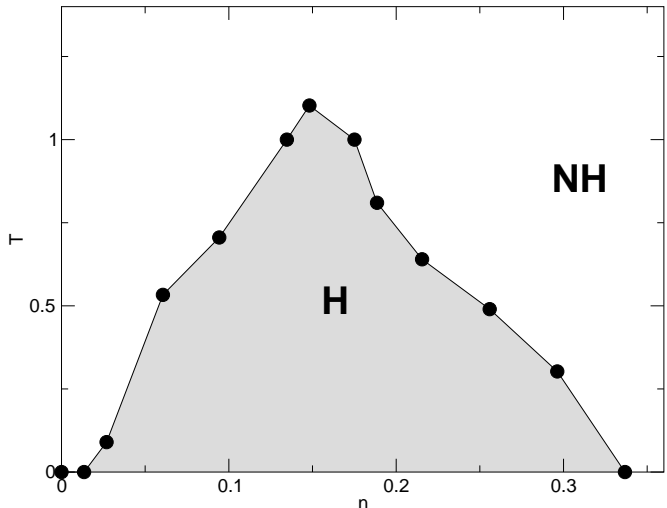}}
\caption{
Temperature $T$ vs density $n$.  The shaded region 
indicates where hysteresis (H) is present
in the velocity-force curve.  The nonhysteretic region is indicated by NH.
}
\end{figure}

\noindent
we have performed a series of simulations for
different densities and temperatures. In Fig.~3 we highlight the region of 
temperature and density where the phases exhibit a finite 
hysteresis in the velocity force curves. 
For high and low densities the transport curves are non-hysteretic at
$T = 0$. As the density increases, the temperature at which 
the hysteresis disappears
increases and reaches a maximum at $n =0.15$, which corresponds to the 
density at which labyrinth or stripe patterns appear. 
As the density is further increased, the
hysteresis width decreases and disappears at $n\approx 0.34$. 
There is some asymmetry in the
hysteretic region, with the hysteresis extending further on the higher
density side of the peak.  This is indicative of the importance of
collective particle interactions in producing the hysteresis; there are
more particles available to interact in the anti-clump state on the high
density side of the peak than in the clump state on the low density side.
In recent magnetization experiments
in doped cuprates, a similar dome-like 
structure was observed where hysteresis is present as 
a function of doping and temperature \cite{Pan29}. A direct comparison 
to the magnetization experiment is difficult since our model does 
not include magnetism; however, 
hysteresis in magnetic materials or superconductors
can be modeled in general 
as collections of particles or domains walls interacting
with quenched disorder and an external driving field, similar to the model
we are using.   
It would be interesting to measure the hysteresis in the 
transport characteristics of the experimental system
as a function of temperature and doping to see if a similar
dome under which hysteresis is present occurs, as found in our simulations. 
We note that hysteresis is not seen in 2D systems with
quenched 
disorder for 
purely repulsive interactions such as vortices \cite{Nori}.   
Hysteresis in transport can appear in 3D vortex systems when the sample breaks
up into two phases and 

\begin{figure}
\center{
\epsfxsize=3.5in
\epsfbox{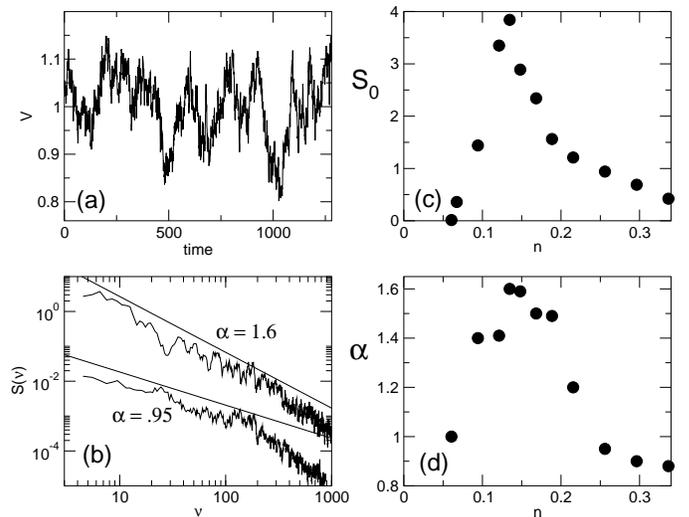}}
\caption{
(a) Time series of the velocity fluctuations for a system
at $T = 0.1$ and $n = 0.15$. (b) The power spectra $S(\nu)$ for the
velocity fluctuations. Upper curve: $n = 0.15$, lower curve: $n=0.3$.
The upper solid line is a power law fit with $\alpha = 1.6$, 
and the lower solid line is a power law fit with $\alpha = 0.95$. 
(c) Noise power $S_{0}$ for a fixed frequency range vs $n$. 
(d) Power spectrum exponent $\alpha$ vs $n$. 
}
\end{figure}

\noindent
these phases form a heterogeneous labyrinth
structure \cite{Dogru}.
This suggests that hysteresis in transport is a sign of
large scale heterogeneities.    

We next examine the noise fluctuations at different densities and 
temperatures.  
We apply a constant drive $f_d$
and measure the velocity fluctuations $\delta V(t)$ as a function of time,
as illustrated in the velocity trace curve of Fig.~4(a)
for a system with $n = 0.15$. From the velocity
fluctuations we can determine the
power spectrum using 
\begin{equation}
S(\nu) = \left|\int \delta V(t) e^{-2\pi i\nu t}dt\right|^2. 
\end{equation}
The noise power $S_{0}$ is defined as the
average value of the power spectrum over a particular frequency octave.       

In Fig.~4(b) we plot the power spectrum for $n = 0.15$, where
the labyrinth phase forms, as well as for $n = 0.3$ where the high density 
uniform phase forms.  Both measurements are performed
at $T = 0.1$.  In the labyrinth phase a $1/f^{\alpha}$ 
noise spectrum is observed with $ \alpha = 1.6$,
while for $n = 0.3$, $\alpha = 1.0$. 
By performing a series of simulations, we can plot the noise power
and $\alpha$ for varied $n$ and $T$. In Fig.~4(c) we show the noise power
$S_0$ as a function of $n$ for fixed $T = 0.1$.
In the hysteretic regions, $\alpha > 1$,
while in the uniform regions, $\alpha \leq 1$. 
The noise power is maximum near $n = 0.15$.  
In Fig.~4(d) we plot $\alpha$, which also shows a maximum in the 
hysteretic regime. 
The power spectra for the Barkhausen type noise seen in the
magnetization experiments were not measured. It 
would be interesting
to examine the change of the noise fluctuations in 
the hysteretic and non-hysteretic regimes. 
Simulations of systems producing Barkhausen noise give

\begin{figure}
\center{
\epsfxsize=3.5in
\epsfbox{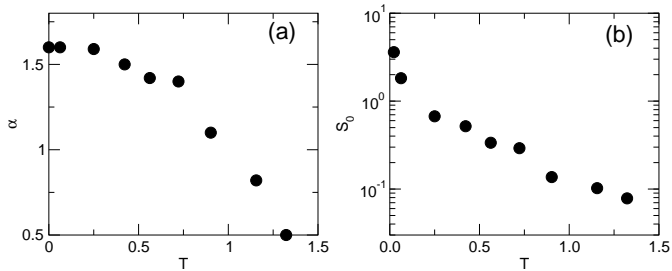}}
\caption{
(a) The exponent $\alpha$ vs $T$ obtained from the power spectrum for
a system with $n = 0.15$. (b) The corresponding noise power $S_{0}$ vs $T$.   
}
\end{figure}

\noindent
$\alpha = 1.77$ \cite{White}.  
The appearance of $\alpha > 1.0$ and large noise power 
in experiments and simulations has   
been interpreted as evidence of glassy dynamics in 
2D charged systems \cite{Popovic,Olson}.   
In contrast, noninteracting or single particle scenarios
for noise generation predict $\alpha \leq 1.0$ 
\cite{Weissman,Weissmanb,Yu}.

We next examine the noise at a fixed density and increasing
temperature as shown in Fig.~5 for
a system with $n = 0.15$. Here the noise power decreases with
increasing temperature and $\alpha$ also decreases, as seen in Fig.~5(a). 
As the temperature increases, the heterogeneities 
begin to melt and the system becomes more uniform. 
In general, in the hysteric regions of the $T$-$n$ phase diagram, 
$\alpha > 1$. In Fig.~5(a) the hysteresis disappears in the
velocity force curves for $T \gtrsim 1.0$. This corresponds
to the temperature
regime where the value of $\alpha$ drops to
$\alpha < 1$. The noise power $S_{0}$ decreases 
exponentially with temperature, as shown in Fig.~5(b).     
Single particle hopping models often predict an
increase in the noise power with temperature rather than the decrease
observed here \cite{Yu}.  This supports the importance of collective
particle interactions in determining the transport properties
and hysteresis.

In summary, we have examined hysteresis and noise in a model system
of heterogeneous charge ordering in the
presence of quenched disorder. We find that at low and high
densities, the system forms non-heterogeneous phases that do not
exhibit hysteresis in the transport curves. At intermediate densities,
however,
the system forms heterogeneous clump and stripe phases which exhibit
hysteresis in  the transport curves. As a function of temperature and
doping, we find a dome region where the hysteresis occurs. This
prediction can be tested in, {\it e.g.}, 
transport measurements in high temperature 
superconductors for varied dopings. We also examine the noise properties
and find that in the heterogeneous and hysteretic regions, the 
noise power is substantially enhanced 
and shows a $1/f^{\alpha}$ noise spectrum characteristic with
$\alpha > 1$. We predict a peak in the noise power and $\alpha$ as
a function of particle density. Additionally both the noise power 
and the exponent $\alpha$ in the heterogeneous regions 
decrease as a function of temperature.
Our results are consistent with recent 
transport and magnetization experiments in cuprate superconductors 
\cite{Bonetti7,Pan8,Pan29}
in which charge heterogeneities are believed to occur.  
 
Acknowledgments---
This work was supported by the US DoE under Contract No.
W-7405-ENG-36.

\end{document}